\documentclass[12pt]{article}

\usepackage{amsmath,amssymb,url,color}
\usepackage{graphicx}
\usepackage{dcolumn}
\usepackage{bm}
\usepackage{color,framed} 

\DeclareMathAlphabet{\mathbi}{OML}{cmm}{b}{it} 
\newcommand{\bel}{\begin{equation}\label}
\newcommand{\ee}{\end{equation}}
\newcommand{\beq}{\begin{eqnarray}\label}
\newcommand{\eeq}{\end{eqnarray}}
\newcommand{\bc}{\begin{center}}
\newcommand{\ec}{\end{center}}

\newcommand{\bk}{\mbox{\boldmath$\hat{k}$}}
\newcommand{\bit}{\begin{itemize}}
\newcommand{\eit}{\end{itemize}}
\newcommand{\ben}{\begin{enumerate}}
\newcommand{\een}{\end{enumerate}}

\newcommand{\varep}{\varepsilon}
\newcommand{\bv}{\mbox{\boldmath$v$}}
\newcommand{\bV}{\mbox{\boldmath$V$}}
\newcommand{\lamres}{\lambda_{\mbox{\small res}}}

\newcommand{\I}{\int_{\Omega}}

\newcommand{\bzeta}{\mbox{\boldmath$\zeta$}}
\newcommand{\bom}{\mbox{\boldmath$\omega$}}
\newcommand{\non}{\nonumber}
\newcommand{\asp}{\alpha_{a}}

\begin{document}

\bc
\textbf{\Large Enstrophy bounds and the range of space-time\\ 
scales in the hydrostatic primitive equations}
\ec
\bc
\textbf{\large J. D. Gibbon and D. D. Holm},
\par\smallskip
Department of Mathematics
Imperial College London\\
London SW7 2AZ, UK
\ec
\noindent
\par\vspace{-3mm}
\begin{abstract}
The hydrostatic primitive equations (HPE) form the basis of most numerical weather, 
climate and global ocean circulation models. Analytical (not statistical) methods 
are used to find a scaling proportional to $\left(Nu\,Ra\,Re\right)^{1/4}$ for 
the range of horizontal spatial sizes in HPE solutions, which is much broader 
than currently achievable computationally.  The range of scales for the HPE is 
determined from an analytical bound on the time-averaged enstrophy of the 
horizontal circulation. This bound allows the formation of very small spatial 
scales, whose existence would excite unphysically large linear oscillation 
frequencies and gravity wave speeds. 
\end{abstract}

\begin{figure}[h!]
\begin{center}
\includegraphics[width=0.5\textwidth]{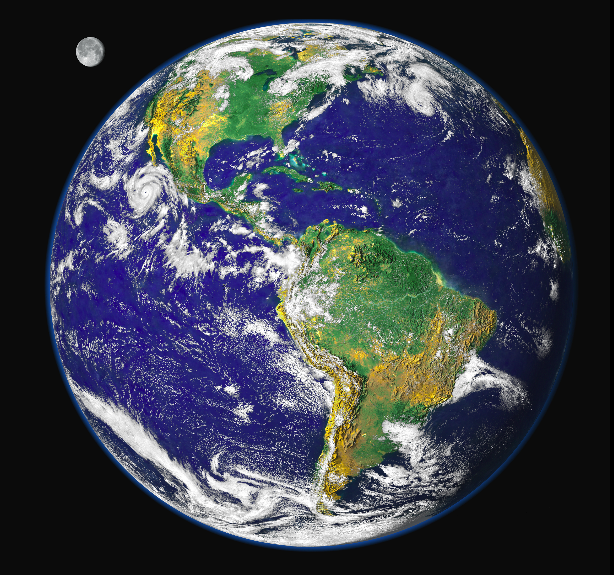}
\caption{\footnotesize A NASA image \cite{NASA} illustrates the  large range of 
fluid scales that exist in atmospheric circulation. The oceanic range of scales 
is similar, but is not so easily observed.}\label{fig1}
\end{center}
\end{figure}

\noindent
The hydrostatic primitive equations (HPE)
have been the foundation of most numerical weather, climate and global ocean 
circulation calculations for many decades 
\cite{Lynchbk,Cullen06,Cullen07,NR02,OhSaMaNa05}. In 
practice, modern computational power can handle integrations of these on global 
horizontal grids ranging in size between 15km and 60km, which correspond respectively 
to one-eighth degree and one-half degree in latitude and longitude at the equator. 
This limitation raises the long-standing question, ``Can numerical simulations 
at these grid sizes adequately predict climate and other natural phenomena that 
occur on the much wider range of scales observed in Nature?" See Figure \ref{fig1}. 
\newpage
Important as it may be, this long-standing question is not addressed here. 
Rather, two questions are addressed associated with the HPE model itself. Namely, ``What 
range of scales is available for solutions of the HPE?'' and ``What scaling law governs 
the size of horizontal HPE excitations in terms of the system parameters?'' These 
dimensionless parameters are $Nu$, $Ra$ and $Re$, associated with the names of 
Nusselt, Rayleigh and Reynolds, respectively. 

The HPE differ from the three-dimensional Navier-Stokes equations in incorporating 
rotation, stratification, and imposing vertical hydrostatic balance. The latter is 
often regarded as the most accurate of the various assumptions used in large-scale 
computations of the climate, weather and ocean  circulation.  The hydrostatic 
assumption determines the pressure from the weight of the water above a given point, 
independently of its state of motion. This changes the nature of the dynamics, because 
the vertical velocity is determined from incompressibility, rather than from its own 
evolution equation. 

Unlike the  Navier-Stokes equations, solutions of the HPE have been proved to be 
regular by Cao and Titi  \cite{CT07}. Moreover, the HPE have also been shown to possess 
a global attractor \cite{NJ07}. Although its solutions are regular, the HPE system may 
potentially possess a vast range of sizes of excitations \cite{GiHo10}. While 
Kolmogorov 
introduced the $Re^{3/4}$ scaling law for the range of spatial sizes of excitations 
in incompressible fluid flows by using statistical methods \cite{Frischbk}, the present 
note will use analytical methods to show that a scaling law exists, proportional 
to $\left(Nu\,Ra\,Re\right)^{1/4}$, for the range of \emph{horizontal} spatial 
sizes in solutions of the HPE, with similar boundary conditions to those of Cao and Titi 
\cite{CT07}. This result demonstrates that HPE excitations are possible at scales that 
are \textit{many} orders of magnitude smaller than are possible in present numerical 
resolutions. 

A dimensionless version of the HPE may be expressed in terms of two sets of velocity vectors 
involving the horizontal velocities $u,\,v$ and the vertical velocity $w$ \cite{Holm96}
\bel{Vvdef}
\bV(x,y,z,t) = (u,\,v,\,\varep w),\qquad\bv = (u,\,v,\,0)
.\ee
Under the constraint of incompressibility, $\mbox{div}\,\bV  = 0$, these satisfy 
\bel{PE1u}
\varep\left(\partial_{t} + \bV\cdot\nabla\right)u - v = \varep Re^{-1}\Delta u - \partial_{x}P
,
\ee
\bel{PE1v}
\varep\left(\partial_{t} + \bV\cdot\nabla\right)v + u = \varep Re^{-1}\Delta v -  \partial_{y}P
.\ee
Here $\varep$ is the Rossby number, $Re = U_{0}L/\nu$ is the Reynolds number and $P$ the pressure. 

As mentioned earlier, HPE has no evolution equation for the vertical velocity component $w$.  
Instead, this variable is determined (diagnosed) from the incompressibility condition, 
$\hbox{div}\,\bV = 0$. The $z$-derivative of the pressure field $P$ and the dimensionless 
temperature $\Theta$ enter through the equation for hydrostatic balance
\bel{hydro1}
a_{0}\Theta +  \partial_{z}P = 0
\,.\ee
The coefficient $a_{0} = (\varep \sigma^{-1}\asp^{-2})R{a}Re^{-2}$ arises from non-dimensionalization 
of the equations. Here $\sigma = \nu/\kappa$ is the Prandtl number (the ratio of viscosity $\nu$ and 
thermal diffusivity $\kappa$), $Ra$ is the Rayleigh number, defined by $R{a} = g\alpha T_{0}H^{3}
(\nu\kappa)^{-1}$, $g$ is  acceleration of gravity, $\alpha$ is volumetric expansion coefficient, 
$T_{0}$ is a typical temperature difference and $\asp = H/L$ is the aspect ratio of the cylindrical 
domain. When (\ref{PE1u}), (\ref{PE1v}) and (\ref{hydro1}) are combined, an evolution equation for 
the hydrostatic velocity field $\bv = (u,\,v,\,0)$ results as
\bel{3DB}
\varep\left(\partial_{t} + \bV\cdot\nabla\right)\bv 
+ \bk\times\bv + a_{0}\bk\Theta = \varep Re^{-1}\Delta\,\bv - \nabla P,
\ee
which is taken in tandem with the incompressibility condition $\hbox{div}\,\bV = 0$. 
The dimensionless temperature $\Theta$ (the source of buoyancy) evolves according to
\bel{tempPDE}
\left(\partial_{t} + \bV\cdot\nabla\right)\Theta = (\sigma Re)^{-1}\Delta\Theta + q
\,,\ee
in which nondimensional $q$ specifies  heat sources, or sinks. The domain $\Omega$ is taken to be 
a cylinder of radius $L$ and height $H$. The vertical 
velocity and vertical flux of horizontal momentum  both vanish on its flat upper and lower cylinder 
surfaces ($z = 0,\,H$). That is, $w = 0$ and $u_{z} = v_{z}= 0$ on the boundary. The variables are 
all taken to be periodic on the sides of the cylinder.

\begin{figure}[htb]
\begin{center}
\includegraphics[width=0.6\textwidth]{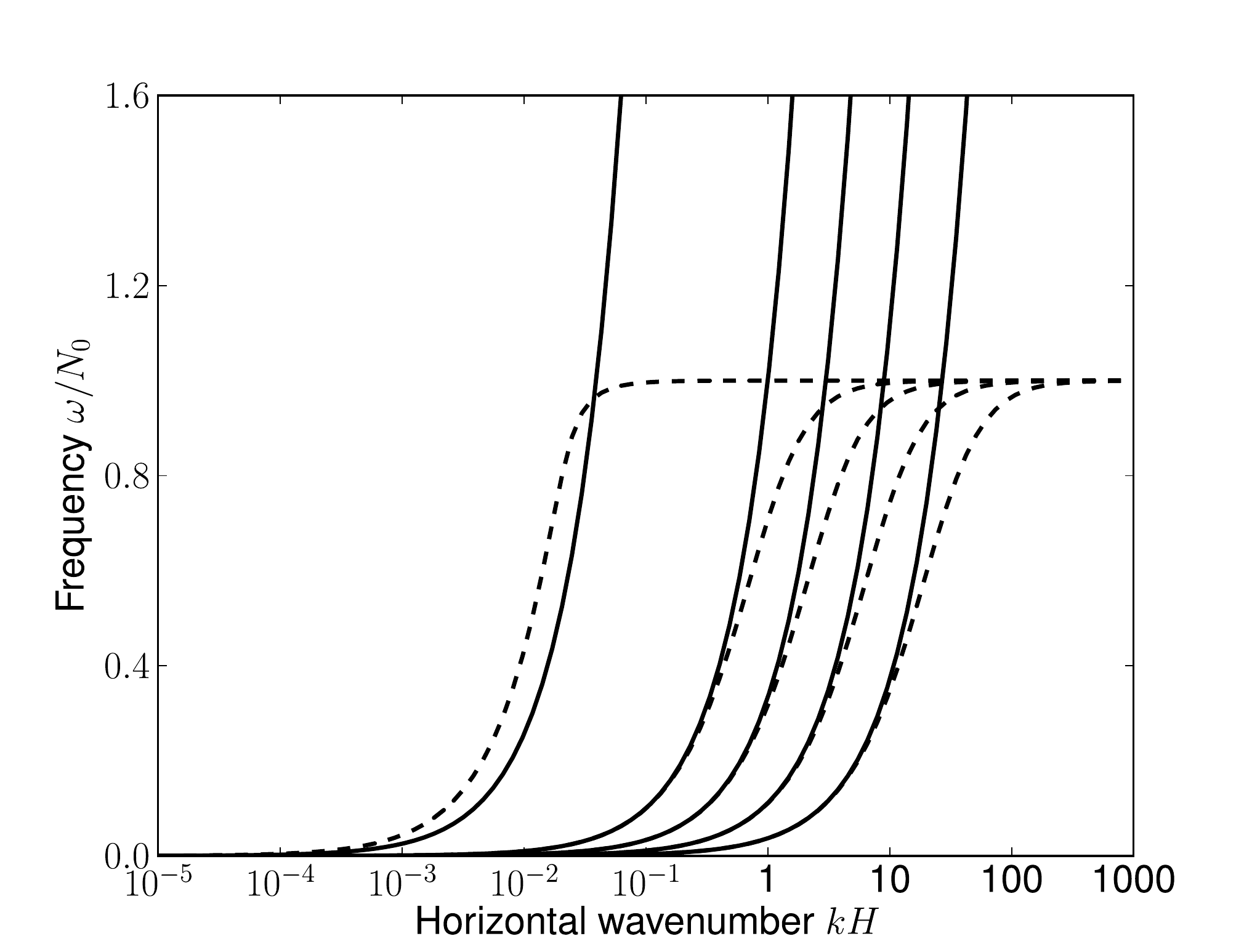}
\caption{\footnotesize This comparison of linear mode dispersion relations for the hydrostatic 
primitive equations (solid curves) with those of the exact nonhydrostatic equations (dashed curves) 
for oceanic conditions shows that the primitive equations admit very high fluctuation frequencies, 
especially at high horizontal wave numbers. In contrast, the dispersion relation for the nonhydrostatic 
equations limits properly to the buoyancy frequency, regardless of how high the horizontal wave number 
becomes. Oceanic parameters are taken as $c_s = 1500 ms^{-1}$, $H = 10^{3.5} m$, $N_0 = 0.01s^{-1}$ 
where the appropriate normalizing length scale $H$ is the mean ocean depth. The multiple curves 
correspond to different choices of vertical wave number, $mH \in [0,1,3,9,27]$, increasing from the left. The value $m=0$ is the barotropic mode and the others are baroclinic.}\label{fig2}
\end{center}
\end{figure}

Linearizing the HPE in (\ref{3DB}) and (\ref{tempPDE}), and their non-hydrostatic equivalent 
(which has the dynamics of $w$ restored), leads to well-known dispersion relations \cite{duk2011}, 
which are illustrated in Figure \ref{fig2}. The essence of these dispersion curves is that without 
the frequency cut-off enforced by the  buoyancy terms in the non-hydrostatic equations, the HPE 
admit unphysically high gravity wave frequencies at small scales. Moreover, these HPE gravity waves 
propagate at a fixed phase speed in the limit of small scales, while in reality  gravity waves at 
these scales cease to propagate at all.

\section{\large An estimate for the resolution length}

Taking the inner product of the divergence-free velocity $\bV$ with the motion 
equation (\ref{3DB}) gives an equation for the rate of change of the  
kinetic energy of horizontal motion
\beq{H1a}
\frac12\frac{d~}{dt} \I |\bv|^{2}\,d\mathcal{V}  =  \I \left(Re^{-1}\bV\cdot\Delta\bv 
- a_{0}w\Theta\right)d\mathcal{V}
\eeq
in which $d\mathcal{V}$ is the volume element and surface terms integrate to zero 
under the present boundary conditions. 
For the Navier-Stokes equations it is normal practice to use the energy dissipation 
rate $\nu\left<\I |\bom|^{2}\,d\mathcal{V}\right>$ based on the full vorticity 
$\bom = \mbox{curl}\,\bV$ to define a length scale called the Kolmogorov length
\cite{DF02}. The quantity $\I |\bom|^{2}\,d\mathcal{V}$ is called the \emph{enstrophy} 
and the angle brackets $\left<\,\cdot\,\right>$ denote the time average
\bel{timavdef}
\left<\,\cdot\,\right> = \lim_{T\to\infty}\frac{1}{T}\int_{0}^{T} (\,\cdot\,)\,dt
\,.\ee
However, it is more appropriate in the hydrostatic approximation to use 
three-dimensional $\bzeta = 
\mbox{curl}\,\bv$ and base a \textit{horizontal} length scale on $\left<\I 
|\bzeta|^{2}\,d\mathcal{V}\right>$, since the vertical velocity $w$ is diagnosed from the 
horizontal velocity dynamics.  To determine this horizontal length scale from the evolution 
of the horizontal kinetic energy in (\ref{H1a}), let us examine the Laplacian term 
\bel{H1b}
\I \bV\cdot\Delta\bv\,d\mathcal{V} = - \I \bom\cdot\bzeta\,d\mathcal{V} 
\ee
where the surface terms again vanish for our choice of boundary conditions. 
Note that $\bzeta$ is fully three dimensional, but its horizontal components vanish 
at the top and bottom of the cylinder.  Two more integrations by parts give
\beq{H1c}
\I \bom\cdot\bzeta\,d\mathcal{V} = \I \left(|\bzeta|^{2} + (\mbox{div}\,\bv)^{2}\right)\,d\mathcal{V}
\geq \I|\bzeta|^{2}\,d\mathcal{V}
\eeq
and thus (\ref{H1a}) may equivalently be re-written as
\bel{H1d}
\frac12 \frac{d~}{dt} \I |\bv|^{2}\,d\mathcal{V} 
\leq - Re^{-1}\!\!\!\I |\bzeta|^{2}\,d\mathcal{V} 
- a_{0}\!\! \I \!w\Theta\,d\mathcal{V}
\,.\ee
Upon defining the vertical Nusselt number $Nu$ as
\bel{Nudef}
Nu := - \left<\I w\Theta\,d\mathcal{V}\right>
,\ee
the time average of (\ref{H1d}) may be written as 
\bel{H1f}
\left<\I |\bzeta|^{2}\,d\mathcal{V}\right> \leq (\varep \sigma^{-1}\asp^{-2})R{a}Re^{-1}Nu
\,,\ee
since the horizontal kinetic energy term vanishes in the limit as $T\to\infty$.
This bound on the time-averaged enstrophy of the horizontal circulation 
$\left<\I |\bzeta|^{2}\,d\mathcal{V}\right>$ yields a \textit{horizontal resolution length 
scale} which emerges upon switching back into dimensional variables. Let $\bzeta_{dim}$ be 
the dimensional version of $\bzeta$\,; that is, $\bzeta_{dim} = L^{-1}U_{0}\bzeta$ for a 
typical horizontal velocity scale $U_{0}$. Then a \emph{resolution scale} $\lamres$ may be 
defined using the same approach as that used to find an analytical estimate of the inverse 
Kolmogorov scale for the Navier-Stokes equations.
\beq{lamdef}
L^{4}\lamres^{-4} &:=& 
L^{4}\left<\left(\nu^{-2}L^{-3}\I |\bzeta_{dim}|^{2}\,d^{3}x\right)\right>\non\\
&=& L^{4}\big(L^{-1}U_{0}\big)^{2}\nu^{-2}\left<\I |\bzeta|^{2}\,d\mathcal{V}\right>\non\\
&=& Re^{2}\left<\I |\bzeta|^{2}\,d\mathcal{V}\right>.
\eeq
Thus, the main result obtained from (\ref{H1f}) and (\ref{lamdef}) is an estimate for the \emph{range} of horizontal scales, defined by the ratio $L\lamres^{-1}$, as 
\bel{lamest}
L\lamres^{-1}\leq \left(\varep \sigma^{-1}\asp^{-2}Nu\,Ra\,Re\right)^{1/4}
.\ee
This bound incorporates all physical processes in their nondimensional forms. 
Estimated from the time-averaged enstrophy of the horizontal circulation, the ratio $L\lamres^{-1}$ of the domain size to the resolution scale provides an upper bound for the range of \emph{horizontal} 
(not vertical) length scales. The hydrostatic approximation holds 
regardless of the magnitude of this ratio.

\section{\large Conclusion}

It is now time to put some numbers into the estimate in (\ref{lamest}). For example, 
in regional flows in the ocean of depth $H\approx 10^{0.5}$km, aspect ratio $\asp 
= 10^{-2}$, Prandtl number $\sigma\approx 10$ and Rossby number $\varep = 10^{-2}$, 
one has $\varep \sigma^{-1}\asp^{-2} \approx 10^1$. Thus, the range of scales 
(\ref{lamest}) in this case may be written as
\bel{thmequn2}
L\lamres^{-1} \lesssim \left(10\,Nu\,Ra\,Re\right)^{1/4}.
\ee
The Rayleigh, Prandtl and Nusselt numbers usually appear in Rayleigh-B\'enard convection 
in which $Nu$ is observed to scale with $Ra$ such that $Nu \sim Ra^{\beta}$ with variations 
around $\beta = 1/3$\,: see \cite{AhGrLo2009} for a discussion of the state of the art for 
heat transfer and large scale dynamics in turbulent Rayleigh-B\'enard convection. However, 
the hydrostatic approximation excludes deep convective processes, in which case $Nu \approx 1$ 
\cite{Hi2011}. The 
Rayleigh-B\'enard $\beta$-scaling for $Nu$ would apply only at small vertical turbulence 
scales where the hydrostatic approximation would be invalid. An important issue in oceanic 
simulations is to differentiate between mass flux and heat flux. Numerical simulations 
of ocean circulation must typically be corrected to prevent over-estimating the heat 
flux \cite{GentMcW}. The need for this correction is another indication that the Nusselt 
number tends to be small in oceanic flows. 
\par\smallskip
The sizes of $Ra$ and $Re$ for typical flows in the ocean are very large, when 
based on regional domain size and molecular values of viscosity and diffusivity 
of heat. For example, with $H \approx 10^{3.5}$m and $Nu \approx 1$ 
\beq{Raest}
R{a} &=& g\alpha T_{0}H^{3}(\nu\kappa)^{-1}\non\\
&\approx&
{10^1}{10^{-4}}{10^0}10^{10.5}({10^6}{10^7}) \approx 10^{20.5}
\eeq
and $Re = U_{0}H/(\nu\alpha_a)\approx 10^{-1}10^{3.5}(10^610^2)\approx 10^{10.5}$. 
According to these estimates, $R{a}Re^{-2} = O(1)$, and the coefficient $a_{0} = 
(\varep \sigma^{-1}\asp^{-2})R{a}Re^{-2}\approx 10^{1}$; so the range of scales is 
bounded by about \emph{eight} orders of magnitude. That is, in this case, 
$L\lamres^{-1} \lesssim 10^{8}$. This means that for a domain size of $400$km 
at a depth of about $4$km, \textit{the horizontal excitation scales could be 
as small as a few millimeters.} In particular, the estimate (\ref{lamest}) 
with $Nu\approx 1$ and $Ra \sim Re^2$  yields
\bel{betaest}
L\lamres^{-1}\leq \left(\varep \sigma^{-1}\asp^{-2}\right)^{1/4}Re^{3/4}
,\ee
which is close to the Kolmogorov range of scales in 3D. The very high linear wave frequencies 
associated with such small horizontal scales would preclude both the physical relevance and 
the computability of the HPE. The conclusion is that improving the resolution of 
HPE numerical solutions may tend to make their results less accurate and much more expensive to perform, because the nonlinear tendency toward much smaller spatial scales produces wave excitations of rapidly increasing linear frequency (as in Fig. 2) that would require reducing the time-step beyond the present limits of computability. Apparently, this fact is already recognized in practice, since the HPE are generally applied to climate simulations, but not to regional simulations.  What this paper shows and emphasizes is that unphysically small spatial scales can potentially be generated in HPE with molecular values for transport coefficients. 
In fact, modulo appropriate adaptations, the same range of scales would be found to hold for the nonhydrostatic equations, although we do not discuss it here because no proof of existence is available for them.

\par\smallskip
Of course, numerical simulations of large-scale circulations in the ocean and atmosphere 
do not use the molecular values of viscosity and diffusivity. Instead, they introduce 
effective values for these quantities due to unresolved scales, associated with turbulent 
`eddies'. These effective values are chosen essentially to make the Reynolds number at the 
horizontal grid scale 
$Re(\Delta x)$ equal to unity. If the scaling $Ra \sim Re^2$ persists for these 
simulations and the Nusselt number at the grid scale is of order unity, then the 
numerical procedure of setting $Re(\Delta x)=1$ might tend to properly resolve the 
hydrostatic excitations of the HPE. However, it may also be good practice in numerical 
simulations using the HPE to  evaluate the dimensionless numbers at the vertical 
grid scale $Nu(\Delta z)$ and $Ra(\Delta z)$ corresponding to the other physical 
aspects of the HPE. Further study of the scaling law $Ra \sim Re^2$ for 
various regimes of ocean and atmosphere circulation might also be fruitful
in determining local values of the ranges of scales. 

\par\medskip\noindent
\textbf{Acknowledgements} We thank J. K. Dukowicz, R. Hide, B. Hoskins, J. C. McWilliams. J. R. Percival
and E. S. Titi for several enlightening conversations. DDH thanks the Royal 
Society for a Wolfson Research Merit Award.


\end{document}